# Hawking Effects on the Entanglement near a Schwarzschild Black Hole

Doyeol Ahn*

*Institute of Quantum Information Processing and Systems, University of Seoul, Seoul 130-743, Republic of Korea*

Hawking radiation effects on an entangled pair near the event horizon of a Schwarzschild black hole are investigated. The Hawking radiation was found to degrade both the quantum coherence of the entangled state and the mutual correlations of the entangled pair. When the black hole evaporated completely, the measure of entanglement vanished, but the classical correlation between the entangled pair still remained.

*To whom correspondence should be addressed.
E-mail: dahn@uos.ac.kr ; davidahn@hitel.net

# I. INTRODUCTION

Some thirty years ago, Hawking discovered that a black hole not only absorbed the matter and radiation but also radiated its energy in the form of black-body radiation and then eventually evaporated [1-3]. This so-called Hawking effect on the information loss in black holes has been a serious challenge to modern physics because it requires a clear understanding of phenomena ranging from gravity to information theory [4-6]. A similar phenomenon, the Unruh effect, was discovered in a more simple case of a uniformly accelerating frame called a Rindler coordinate [7-11]. From a mathematical point of view, both the Hawking and the Unruh effects arise from the logarithmic phase singularity of wave functions [12], and these effects are expected to occur for sound waves from the sonic horizon in transonic fluid flow [13], electromagnetic waves in matter at a surface of singular electric and magnetic permeabilities [14], dilute-gas Bose-Einstein condensates in the hydrodynamic limit [15], and slow light in an atomic medium [12]. At this stage, it seems to be an interesting study to consider the Hawking or Unruh effect on the quantum entangled states that are essential resources of quantum information processing because understanding entanglement in a relativistic region is important not only for logical completeness but also for the study of the physical bounds of quantum information processing tasks [16-18]. Recently, the Unruh effect on an entangled pair in a Rindler coordinate was studied [19-21], and Unruh effect was found to act as a decoherence process [22,23], so the quantum entanglement is degraded in Rindler frame.

In this paper, the author studied the Hawking effect on the entangled state near the event horizon of a Schwarzschild black hole. The Hawking effect arises from the logarithmic phase singularity in the wave functions in the Schwarzschild frame and a Bogoliubov transformation between Kruskal and Schwarzschild frames. The Hawking radiation was found to degrade both the quantum coherence of the entangled state and the mutual correlations of the entangled pair. When a black hole evaporated completely, the measure of entanglement was found to vanish, but the classical correlations between the entangled pair still remained.

# II. HAWKING EFFECT ON THE ENTANGLED PAIR NEAR THE EVENT HORIZON OF A SCHWARZSCHILD BLACK HOLE

## 1. Hawking Radiation

In this section, we first review the derivation of Hawking radiation [6, 7] and extend the results to an entangled state near the event horizon. The stationary Schwarzschild black hole is represented by the metric

$$ds^2 = -\left(1-\frac{2M}{r}\right)dt^2 + \frac{dr^2}{1-\frac{2M}{r}} + r^2\left(d\theta^2 + \sin^2\theta d\varphi^2\right), \tag{1}$$

where $M$ is the mass of the black hole. At $r = 2M$, the Schwarzschild spacetime has an event horizon. The general coordinate is $x^\mu = (t, r, \theta, \varphi)$ with the metric tensor given by

$$g_{tt} = -\left(1-\frac{2M}{r}\right),\ g_{rr} = \frac{1}{1-\frac{2M}{r}},\ g_{\theta\theta} = r^2,\ g_{\varphi\varphi} = r^2\sin^2\theta\ . \tag{2}$$

The massless scalar field satisfies the wave equation

$$(-g)^{1/2}\frac{\partial}{\partial x^\mu}\left[g^{\mu\nu}(-g)^{1/2}\frac{\partial}{\partial x^\nu}\right]\phi = 0, \tag{3}$$

and the positive-frequency normal-mode solution is given by [6]

$$\phi_{\omega lm} = (2\pi|\omega|)^{-1/2}e^{-i\omega t}f_{\omega l}(r)Y_{lm}(\theta,\varphi), \tag{4}$$

where $f_{\omega l}(r)$ satisfies

$$\frac{\partial^2 f_{\omega l}}{\partial r^{*2}} + \omega^2 f_{\omega l} - \left(1-\frac{2M}{r}\right)\left(\frac{l(l+1)}{r^2} + \frac{2M}{r^3}\right)f_{\omega l} = 0, \tag{5}$$

with $r^* = r + 2M\ln(r/2M - 1)$. If we denote the radiation part of the wave coming out of the past horizon of the black hole by $f_{\omega l}^-$, then it is given by

$$f_{\omega l}^-(r) \approx e^{i\omega r^*} + A_{\omega l}^- e^{-i\omega r^*}. \tag{6}$$

In Kruskal coordinates, the Schwarzschild metric becomes [6, 9]

$$\begin{aligned}
&ds^2 = -2M\frac{e^{-r/2M}}{r}d\bar{u}d\bar{v} + r^2 d\theta^2 + r^2\sin^2\theta d\varphi^2,\\
&\bar{u} = -4Me^{-u/4M},\ \bar{v} = 4Me^{v/4M},\\
&u = t - r^*,\ v = t + r^*,\\
&r^* = r + 2M\ln(r/2M - 1).
\end{aligned} \tag{7}$$

The Kruskal extension of the Schwarzschild spacetime is shown in Fig. 1. Since the Killing vector in Kruskal coordinates is given by $\partial/\partial \bar{u}$ on $H^-$ (Fig. 1), the solution in Kruskal coordinates is given by

$$\bar{\phi}_{\varpi lm} = (2\pi |\varpi|)^{-1/2} e^{-i\varpi \bar{u}} Y_{lm}(\theta, \varphi). \tag{8}$$

On the other hand, the original positive-frequency normal-mode on $H^-$ can be written as

$$\phi^-_{\omega lm} = (2\pi |\omega|)^{-1/2} (e^{-i\omega u} + A^-_{\omega l} e^{-i\omega v}) Y_{lm}(\theta, \varphi). \tag{9}$$

Using $e^{-i\omega u} = (|\bar{u}|/4M)^{i4M\omega}$ and $e^{-i\omega v} = (\bar{v}/4M)^{-i4M\omega}$ and the fact that $\bar{v} = 0$ on $H^-$ [9], we obtain

$$\phi^-_{\omega lm} = (2\pi |\omega|)^{-1/2} (|\bar{u}|/4M)^{i4M\omega} Y_{lm}(\theta, \varphi). \tag{10}$$

Since, $\bar{u} < 0$ in region $I$ and $\bar{u} > 0$ in region $II$ of Fig. 1, the wave coming out of the past horizon of the black hole on $H^-$ can be written as

$$\phi^-_{\omega lm} = (e^{2\pi M\omega} {}_{out}\phi_{\omega lm} + e^{-2\pi M\omega} {}_{in}\phi_{\omega lm})/(2\sinh(4\pi M\omega))^{1/2}, \tag{11}$$

where ${}_{out}\phi_{\omega lm}$ vanishes inside the event horizon (region $II$) and ${}_{in}\phi_{\omega lm}$ vanishes in the exterior region of the black hole (region $I$). In Eq. (11), we have used the fact that $(-1)^{-i4M\omega} = e^{4\pi M\omega}$.

The above definition of the positive-frequency solution in terms of ${}_{out}\phi_{\omega lm}$ and ${}_{in}\phi_{\omega lm}$ leads to the Bogoliubov transformations [9-11] for the particle creation and annihilation operators in the Schwarzschild and the Kruskal spacetimes:

$$a_{K,\omega lm} = \cosh r_\omega b_{out,\omega lm} - \sinh r_\omega b^\dagger_{in,\omega lm},$$
$$a^\dagger_{K,\omega lm} = \cosh r_\omega b^\dagger_{out,\omega lm} - \sinh r_\omega b_{in,\omega lm}, \tag{12}$$
$$\tanh r_\omega = e^{-4\pi M\omega}, \quad \cosh r_\omega = (1 - e^{-4\pi M\omega})^{-1/2},$$

where $a^\dagger_{K,\omega lm}$ and $a_{K,\omega lm}$ are, respectively, the creation and the annihilation operators acting on the Kruskal vacuum in region $I$, $b^\dagger_{out,\omega lm}$ and $b_{out,\omega lm}$ are the creation and the annihilation operators acting on the Schwarzschild vacuum of the exterior region of a black hole, and $b^\dagger_{in,\omega lm}$ and $b_{in,\omega lm}$ are the creation and the annihilation operators acting on the Schwarzschild vacuum inside the event horizon. Then, the ground state in the Kruskal spacetime for mode $(\omega lm)$ is given by $|0_\omega\rangle_K = |\Phi_o(\omega lm)\rangle_{in \otimes out}$, which looks like the vacuum in the far past is a maximally entangled two-mode squeezed state on $H_{in} \otimes H_{out}$ [9, 11, 24]:

$$|\Phi_o(\omega lm)\rangle_{in \otimes out} = \frac{1}{\cosh r_\omega} \sum_n e^{-4\pi M\omega n} |n(\omega lm)\rangle_{in} \otimes |n(\omega lm)\rangle_{out}, \tag{13}$$

where $\{|n(\omega lm)\rangle_{in}\}$ and $\{|n(\omega lm)\rangle_{out}\}$ are orthonormal bases (normal-mode solutions) for mode $(\omega lm)$ for $H_{in}$ and $H_{out}$, respectively. Equation (13) shows that the original vacuum state evolves to a two-mode squeezed state $|\Phi_o\rangle_{in\otimes out}$, which is also called the Unruh state [8-9] and resides on $H_{in} \otimes H_{out}$. This state contains a flux of outgoing particles in $H_{out}$. For an observer outside the black hole, the unitarity is lost because he will not be able to make any measurement in $H_{in}$; as a result, he will be forced to take an average over the states in $H_{in}$ to obtain the density operator in $H_{out}$.

## 2. Entangled Pair near the Horizon

Now, we consider the maximally entangled quantum state shared by Alice and Bob consisting of two modes $(\omega lm)$ and $(\omega' l' m')$ of a free massless scalar field in the Kruskal spacetime, which is given by

$$|\Psi\rangle_K^{AB} = \frac{1}{\sqrt{2}}\left(|0_\omega\rangle_K^A \otimes |0_{\omega'}\rangle_K^B + |1_\omega\rangle_K^A \otimes |1_{\omega'}\rangle_K^B\right), \qquad (14)$$

where

$$\begin{aligned}
|0_\omega\rangle_K^A &= \frac{1}{\cosh r_\omega} \sum_n e^{-4\pi M \omega n} |n(\omega)\rangle_{in} \otimes |n(\omega)\rangle_{out}, \\
|1_\omega\rangle_K^A &= \frac{1}{\cosh^2 r_\omega} \sum_n e^{-4\pi M \omega n} \sqrt{n(\omega)+1} |n(\omega)\rangle_{in} \otimes |n(\omega)+1\rangle_{out},
\end{aligned} \qquad (15)$$

and

$$\begin{aligned}
|0_{\omega'}\rangle_K^B &= \frac{1}{\cosh r_{\omega'}} \sum_n e^{-4\pi M \omega' n} |n(\omega')\rangle_{in} \otimes |n(\omega')\rangle_{out}, \\
|1_{\omega'}\rangle_K^A &= \frac{1}{\cosh^2 r_{\omega'}} \sum_n e^{-4\pi M \omega' n} \sqrt{n(\omega')+1} |n(\omega')\rangle_{in} \otimes |n(\omega')+1\rangle_{out}.
\end{aligned} \qquad (16)$$

In Eqs. (15) and (16), we used the relation $|1_\omega\rangle_K = a^\dagger_{K,\omega lm} |0_\omega\rangle_K$ and represented the mode by a single index $\omega$ to simplify the notation. Observers such as Alice and Bob outside the event horizon of the black hole are causally disconnected from the interior of the horizon, so the density operator for the outside of the black hole is given by

$$\begin{aligned}
\rho^{AB} &= Tr_{in}\left(|\Psi\rangle_K^{AB}\langle\Psi|\right) \\
&= \frac{1}{\cosh^2 r_\omega \cosh^2 r_{\omega'}} \sum_{n,q} \tanh^{2n} r_\omega \tanh^{2q} r_{\omega'} \rho^{AB}_{nq},
\end{aligned} \qquad (17)$$

where

$$\rho_{nq}^{AB} = \frac{1}{2}|nq\rangle_{\omega\omega'}\langle nq|$$
$$+ \frac{\sqrt{(n(\omega)+1)(q(\omega')+1)}}{2\cosh r_\omega \cosh r_{\omega'}}\left(|(n+1)(q+1)\rangle_{\omega\omega'}\langle nq| + |nq\rangle_{\omega\omega'}\langle(n+1)(q+1)|\right) \quad (18)$$
$$+ \frac{(n(\omega)+1)(q(\omega')+1)}{2\cosh^2 r_\omega \cosh^2 r_{\omega'}}|(n+1)(q+1)\rangle_{\omega\omega'}\langle(n+1)(q+1)|,$$

with $|nq\rangle_{\omega\omega'} = |n(\omega)\rangle_{out} \otimes |q(\omega')\rangle_{out}$.

In this work, the author is interested in how the entanglement near the event horizon is changed by the Hawking radiation. The measure of entanglement can be found by using the partial transpose criterion [25, 26]. If at least one eigenvalue of the partial transpose is negative, the density matrix is entangled. The matrix representation of $\rho_{nq}^{AB}$ in the [nq, n(q+1), (n+1)q, (n+1)(q+1)] block is given by

$$\rho_{nq}^{AB} = \begin{pmatrix} 1/2 & 0 & 0 & a/2 \\ 0 & 0 & 0 & 0 \\ 0 & 0 & 0 & 0 \\ a/2 & 0 & 0 & a^2/2 \end{pmatrix}, \quad (19)$$

where $a = \frac{\sqrt{(n(\omega)+1)(q(\omega')+1)}}{\cosh r_\omega \cosh r_{\omega'}}$. Then, the eigenvalues of the partial transposed density matrix are obtained as

$$\lambda = \frac{1}{2}, \; -\frac{\sqrt{(n(\omega)+1)(q(\omega')+1)}}{2\cosh r_\omega \cosh r_{\omega'}}, \; \frac{\sqrt{(n(\omega)+1)(q(\omega')+1)}}{2\cosh r_\omega \cosh r_{\omega'}}, \; \frac{(n(\omega)+1)(q(\omega')+1)}{2\cosh^2 r_\omega \cosh^2 r_{\omega'}}. \quad (20)$$

The above result shows that one eigenvalue is always negative but approaches zero in the limit $M \to 0$ (or $\cosh r_\omega, \cosh r_{\omega'} \to \infty$), which corresponds to the evaporation of the black hole.

In Fig. 2, we plot the entanglement measure $E_N = 2|\lambda_-|$ as a function of the squeezing parameter $r_\omega = \tanh^{-1}[\exp(-4\pi M\omega)]$. Here, $\lambda_-$ is the negative eigenvalue of the partial-transposed density matrix, and we assumed $\omega = \omega'$ to simplify the computation. The monotonous decrease of $E_N$ with increasing $r_\omega$ indicates that the quantum coherence of the initial entangled pair is lost to the thermal fields generated by the Hawking effects. This result agrees well with Hawking's original argument [2], which says that smaller black holes are at a higher temperature and, thus, radiate more violently than massive black holes. In essence, the Hawking fields act as heat baths for the initial entangled pair.

It is also interesting to estimate the total amount of correlation in the entangled pair by calculating the mutual information as a function of $r_\omega$. The mutual information is defined by [19, 27-29]

$$I(\rho^{AB}) = S(\rho^A) + S(\rho^B) - S(\rho^{AB}), \qquad (21)$$

where $S(\rho)$ is the von Neumann entropy. After some mathematical manipulations, one obtains

$$S(\rho^{AB}) = -\sum_{n,q} \frac{\tanh^{2n} r_\omega \tanh^{2q} r_{\omega'}}{2\cosh^2 r_\omega \cosh^2 r_{\omega'}} \left(1 + \frac{(n+1)(q+1)}{\cosh^2 r_\omega \cosh^2 r_{\omega'}}\right) \\ \times \log_2 \left\{\frac{\tanh^{2n} r_\omega \tanh^{2q} r_{\omega'}}{2\cosh^2 r_\omega \cosh^2 r_{\omega'}} \left(1 + \frac{(n+1)(q+1)}{\cosh^2 r_\omega \cosh^2 r_{\omega'}}\right)\right\}, \qquad (22)$$

$$S(\rho^A) = 1 - \frac{1}{2}\sum_n \frac{\tanh^{2n} r_\omega}{\cosh^2 r_\omega} \log_2 \frac{\tanh^{2n} r_\omega}{\cosh^2 r_\omega} \\ - \frac{1}{2}\sum_n \frac{(n+1)\tanh^{2n} r_\omega}{\cosh^4 r_\omega} \log_2 \frac{(n+1)\tanh^{2n} r_\omega}{\cosh^4 r_\omega}, \qquad (23)$$

and

$$I(\rho^{AB}) = 1 - \frac{1}{2}\sum_n \frac{\tanh^{2n} r_\omega}{\cosh^2 r_\omega} \log_2 \frac{\tanh^{2n} r_\omega}{\cosh^2 r_\omega} \\ - \frac{1}{2}\sum_n \frac{(n+1)\tanh^{2n} r_\omega}{\cosh^4 r_\omega} \log_2 \frac{(n+1)\tanh^{2n} r_\omega}{\cosh^4 r_\omega} \\ + 1 - \frac{1}{2}\sum_q \frac{\tanh^{2q} r_\omega}{\cosh^2 r_\omega} \log_2 \frac{\tanh^{2q} r_\omega}{\cosh^2 r_\omega} \\ - \frac{1}{2}\sum_q \frac{(n+1)\tanh^{2q} r_\omega}{\cosh^4 r_\omega} \log_2 \frac{(q+1)\tanh^{2q} r_\omega}{\cosh^4 r_\omega} \\ + \sum_{n,q} \frac{\tanh^{2n} r_\omega \tanh^{2q} r_\omega}{2\cosh^2 r_\omega \cosh^2 r_\omega} \left(1 + \frac{(n+1)(q+1)}{\cosh^2 r_\omega \cosh^2 r_\omega}\right) \\ \times \log_2 \left\{\frac{\tanh^{2n} r_\omega \tanh^{2q} r_\omega}{2\cosh^2 r_\omega \cosh^2 r_\omega} \left(1 + \frac{(n+1)(q+1)}{\cosh^2 r_\omega \cosh^2 r_\omega}\right)\right\}. \qquad (24)$$

Figure 3 shows $S(\rho^{AB})$, $S(\rho^A)$, and $I(\rho^{AB})$ as functions of $r_\omega$. We also assumed $\omega = \omega'$. Initially, the mutual information has a value of 2. As the mass of the black hole decreases, the mutual information becomes smaller, converging to unity when the black hole finally evaporates ($M \to 0$). The entropy for the joint density matrix $\rho^{AB}$ is zero initially because it's the pure state, has a maximum uncertainty around $r_\omega = 2$, and then approaches unity, which gives the classical correlation of maximally mixed states when the black hole evaporates. The change in the entropy is due to a coupling with the Hawking radiation, i.e., an entanglement of modes in $H_{out}$ with modes in $H_{in}$. In this work, the effect of the final state boundary condition [4-6] on the entanglement at the event horizon is not considered and is left for a future work.

# III SUMMARY

In summary, the Hawking effect on the entangled state near the even horizon of a Schwarzschild black hole is studied. The Hawking effect arises from a logarithmic phase singularity in the wave functions in a Schwarzschild frame and a Bogoliubov transformation between the Kruskal and the Schwarzschild frames. The Hawking radiation was found to degrade both the quantum coherence of the entangled state and the mutual correlations of the entangled pair. When the black hole evaporated completely, the measure of entanglement vanished but the classical correlations between the entangled pair remained.

# ACKNOWLEDGMENT

This work was supported by the University of Seoul through a University Research Grant for 2006.

**Figure legends**

**Fig. 1.** Kruskal extention of the Schwarzschild spacetime [6, 8]. In region *I*, null asymptotes $H_+$ and $H_-$ act as futue and past event horizons, respectively. The boundary lines labelled $J^+$ and $J^-$ are future and past null infinities, respectively, and $i^o$ is the spacelike infinity.

**Fig. 2.** Measure of the entanglement $E_N = 2|\lambda_-|$ as a function of the squeezing parameter $r_\omega = \tanh^{-1}[\exp(-4\pi M\omega)]$. Here, $\lambda_-$ is the negative eigenvalue of the partial-transposed density matrix, and we assumed $\omega = \omega'$ to simplify the computation.

**Fig. 3.** von Neumann entropies $S(\rho^{AB})$ and $S(\rho^A)$, and mutual information $I(\rho^{AB})$ as functions of the squeezing parameter $r_\omega = \tanh^{-1}[\exp(-4\pi M\omega)]$.

Fig. 1

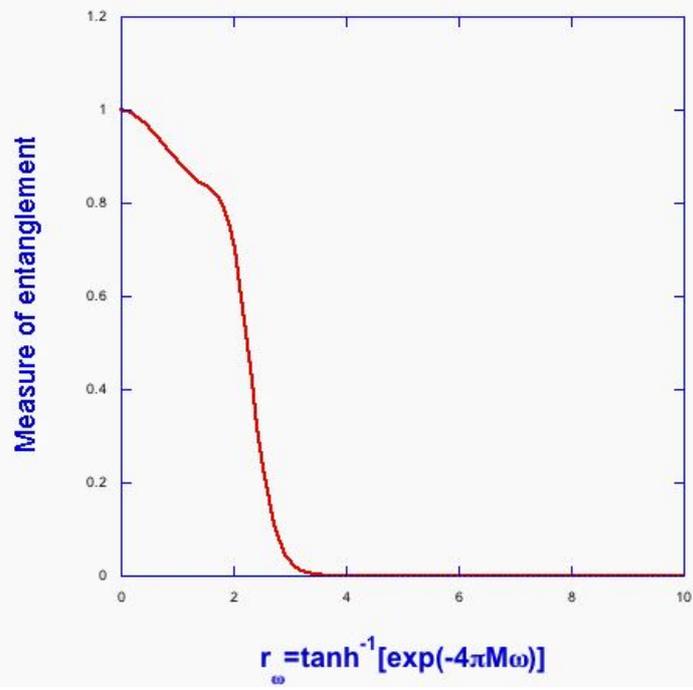

Fig. 2

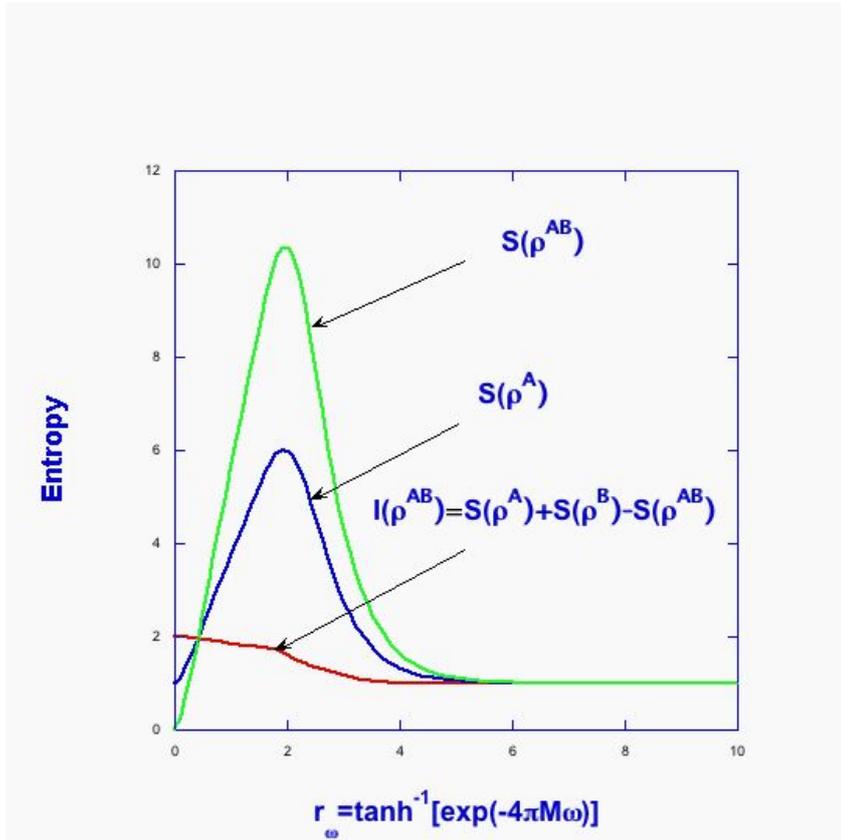

Fig. 3